\documentclass[twocolumn,showpacs,preprintnumbers,amsmath,amssymb,unsortedaddress]{revtex4}

\usepackage{times}
\usepackage{graphicx}
\usepackage{dcolumn}
\usepackage{bm}
\usepackage{amsmath}
\def\eps{\varepsilon}

\begin{document}
\title{Theory of incompressible MHD turbulence with scale-dependent alignment and cross-helicity}
\author{J.~J.~Podesta and A.~Bhattacharjee}
\affiliation{%
Center for Integrated Computation and Analysis of Reconnection and Turbulence,%
University of New Hampshire, Durham, New Hampshire, 03824}%




\date{\today}
\begin{abstract}
A phenomenological anisotropic theory of  MHD turbulence with nonvanishing cross-helicity is constructed based 
on Boldyrev's phenomenology and probabilities $p$ and $q$ for fluctuations 
$\delta \bm v_\perp$ and $\delta \bm b_\perp$ to be positively or negatively aligned.
The positively aligned fluctuations occupy a fractional
volume $p$ and the negatively aligned fluctuations occupy a fractional volume $q$.
Guided by observations suggesting that the normalized cross-helicity $\sigma_c$ and the
probabilities $p$ and $q$ are 
approximately scale invariant in the inertial range, a generalization of Boldyrev's theory is 
derived that depends on the three ratios $w^+/w^-$, $\epsilon^+/\epsilon^-$, and $p/q$.
It is assumed that the cascade process for positively and negatively aligned fluctuations 
are both in a state of critical balance and that the eddy geometries are scale invariant.  
The theory reduces to Boldyrev's original theory when $\sigma_c=0$, 
$\epsilon^+ =\epsilon^-$, and $p=q$ and 
extends the theory of Perez and Boldyrev when $\sigma_c\ne 0$.  
The theory is also an anisotropic generalization of the theory
of Dobrowolny, Mangeney, and Veltri.
\end{abstract}

\maketitle

\section{Introduction}

Phenomenological theories of incompressible MHD turbulence
that take into account the anisotropy of the fluctuations with respect
to the direction of the mean magnetic field $\bm B_0$ were pioneered by
Goldreich \& Sridhar and others in the 1990s.  The influencial and
somewhat controversial theory of Goldreich \& Sridhar 
\citep{Goldreich_Sridhar:1995,Goldreich_Sridhar:1997} established the idea that 
the timescale or coherence time for motions of a turbulent eddy parallel 
and perpendicular to $\bm B_0$ must be equal
to each other and this unique timescale then defines the energy cascade time.
This concept, called {\it critical balance}, leads immediately to the 
perpendicular energy spectrum $E(k_\perp)\propto k_\perp^{-5/3}$ and to the
scaling relation $k_\parallel \propto k_\perp^{2/3}$ describing the anisotropy of the
turbulent eddies.  
\medskip

The decade following the publication 
of the paper by Goldreich \& Sridhar in 1995 was a time when
significant advances in computing power were brought to bear on computational studies
of MHD turbulence.  Simulations of
incompressible MHD turbulence during this time showed that when the
mean magnetic field is strong, $B_0^2 \gg (\delta b)^2$, the perpendicular 
energy spectrum exhibits a power-law scaling closer to 
$k_\perp^{-3/2}$ than $k_\perp^{-5/3}$ \citep{Maron:2001,Muller:2003,Muller:2005}.  
Motivated by this result, Boldyrev modified the Goldreich \& Sridhar theory
to explain the $k_\perp^{-3/2}$ power-law seen in simulations \citep{Boldyrev:2005,
Boldyrev:2006}.  A new concept that emerged in Boldyrev's theory is the scale 
dependent alignment of velocity and magnetic field fluctuations whereby the 
angle $\theta$ formed by $\delta \bm v_\perp$ and $\delta \bm b_\perp$
scales like $\lambda_\perp^{1/4}$ in the inertial range.  This alignment effect weakens the
nonlinear interactions and yields the perpendicular energy spectrum
$E(k_\perp)\propto k_\perp^{-3/2}$.  Evidence for this alignment effect
has been found in numerical simulations of incompressible MHD turbulence
\citep{Mason:2006,Mason:2008} 
and in studies of solar wind data \citep{Podesta_Bhattacharjee:2008,
Podesta_Chandran:2009}.  
\medskip

The phenomenological theory of 
\citet{Galtier:2005} can also be used to explain the observed $k_\perp^{-3/2}$ 
energy spectrum.  Using a slightly modified critical balance relation that retains the 
$k_\parallel \propto k_\perp^{2/3}$ scaling of the Goldreich \& Sridhar
theory, their model admits the $k_\perp^{-3/2}$ energy spectrum (as well as
other solutions).  However, the theory of \citet{Galtier:2005} does not include the 
scale dependent alignment that arises in Boldyrev's theory and, more importantly, 
is seen in the solar wind \citep{Podesta_Bhattacharjee:2008,Podesta_Chandran:2009}.
\medskip

The theories discussed so far \citep{Goldreich_Sridhar:1995,
Goldreich_Sridhar:1997,Boldyrev:2005,Boldyrev:2006,Galtier:2005} all assume that the 
cross-helicity vanishes and, therefore, these theories cannot be applied 
to solar wind turbulence.  When the cross-helicity of the turbulence is
nonzero it is necessary to take into
account the cascades of both energy and cross-helicity.
A generalization of the Goldreich \& Sridhar theory to turbulence with
nonvanishing cross-helicity, also called imbalanced turbulence, has been developed
by Lithwick, Goldreich, \& Sridhar \citep{Lithwick_Goldreich:2007}.
Other theories of imbalanced turbulence have been derived by Beresnyak \& Lazarian
\citep{Beresnyak_Lazarian:2008} and Chandran \citep{Chandran:2008,Chandran:2009}.
However, none of these theories contains the scale dependent alignment
of velocity and magnetic field fluctuations seen in the solar wind.
Therefore, to develop a theory that may be applicable to solar wind turbulence it 
is of interest to generalize Boldyrev's theory to incompressible MHD 
turbulence with non-vanishing cross-helicity.  An extension of Boldyrev's theory 
to imbalanced turbulence has been discussed by Perez \& Boldyrev
\citep{Perez:2009}.  The purpose of the present paper is to develop a  
theoretical framework which generalizes the results of Perez \& Boldyrev
and is consistent with solar wind observations.
\medskip

Our theory is founded, in part, on two new solar wind observations presented in this paper.  
The first is the observation that the normalized cross-helicity $\sigma_c$, the ratio of 
cross-helicity to energy, is \emph{scale invariant} in the inertial range.
The second is the observation that the probabilities $p$ and $q$  that fluctuations are
positively or negatively aligned, respectively,  are also \emph{scale invariant}, 
that is, these quantities are approximately constant in the inertial range.  
Experimental evidence 
for the scale invariance of $\sigma_c$ comes from solar wind observations 
by \citet{Marsch_Tu:1990} and also Figure~1 below, and from numerical simulations 
\citep{Verma_Roberts:1996,Perez:2009}. Evidence for the scale invariance of $p$ and $q$ is 
shown in Figure~2 below.  Assuming these quantities are all scale invariant we deduce 
expressions for the energy cascade rates and the rms fluctuations that generalize 
the results in \citep{Boldyrev:2006} and \citep{Perez:2009} and are consistent with 
the concept of scale dependent alignment 
of velocity and magnetic field fluctuations, a concept neglected in other phenomenological 
theories \citep{Galtier:2005,Lithwick_Goldreich:2007,Beresnyak_Lazarian:2008,Chandran:2008}.  
The resulting theory, which is founded on the concept of scale-invariance and grounded 
in solar wind observations, contains the theories of \citet{Boldyrev:2006} and
\citet{Perez:2009} as special cases, but opens up a broader range of physical possibilities.
\medskip

Consistent with numerical simulations and solar wind observations, in our approach
the fluctuations at a given point may assume one of two possible states referred to
as {\it positively aligned} $\delta \bm v_\perp \cdot \delta \bm b_\perp >0$
and {\it negatively aligned} $\delta \bm v_\perp \cdot \delta \bm b_\perp <0$.
Each state is characterized by its own rms energy $v^2$, alignment angle $\theta$,
and nonlinear timescale $\tau$.  Positively aligned fluctuations have a characteristic 
spatial gradient which determines their nonlinear timescale and 
negatively aligned fluctuations have a different 
spatial gradient which determines their nonlinear timescale.
These timescales are estimated from the nonlinear terms in the MHD equations
as described in sections 2 and 3.

Section 2 describes the geometries of velocity and magnetic field fluctuations that are either 
aligned `$\uparrow$'
or anti-aligned `$\downarrow$' and these are used to form estimates of the nonlinear terms in the
MHD equations.  From this foundation, estimates of the energy cascade times 
are constructed in section 3 and the theory of the energy cascade process is developed in 
section 4.  The summary and conclusions are presented in section 5.

\section{Fluctuations in imbalanced turbulence}

Consider velocity and magnetic field fluctuations measured between two points
separated by a distance $\lambda_\perp$ in the field perpendicular plane.  
Let $\bm v$ and $\bm b$ denote the fluctuations in the plane
perpendicular to the local mean magnetic field, where $\bm v$ and $\bm b$ 
are both measured in velocity units.  Suppose that $\bm v$ and $\bm b$ are aligned with
some small angle $\theta>0$ and assume, as for Alfv\'en waves, that $|\bm v|=|\bm b|$.  
Then $\bm w^+=\bm v+\bm b$ is nearly aligned
with $\bm v$ and $\bm w^-=\bm v-\bm b$ is nearly perpendicular to $\bm v$ as sketched 
in Figure 1.  
\begin{figure*}
\includegraphics[width=4.5in]{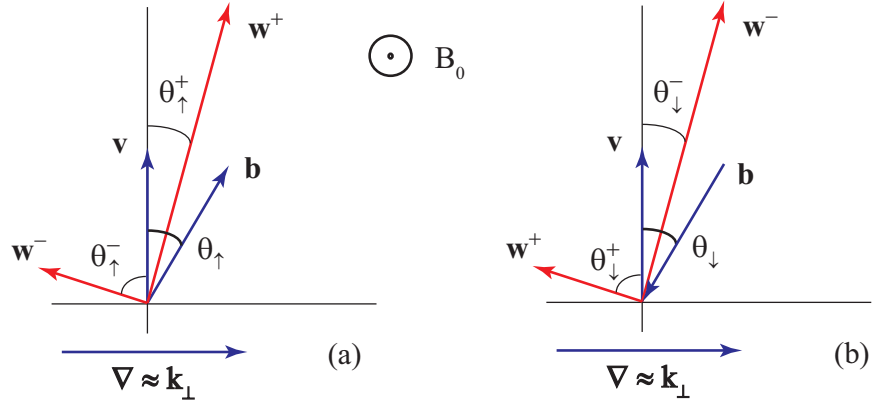}%
\caption{\label{fig0}%
     Geometry of the fluctuation vectors $\bm v$ and $\bm b$ for positively aligned
fluctuations (a) denoted by `$\uparrow$' and negatively aligned fluctuations (b)
denoted by `$\downarrow$'.  The gradient is perpendicular to the velocity fluctuation $\bm v$.  
The magnitude of $\bm v$ for positively and negatively aligned fluctuations
are $v_\uparrow$ and $v_\downarrow$, respectively.  The angles formed by $\bm v$ and $\bm b$,
$\theta_\uparrow$ and $\theta_\downarrow$, are both assumed to be small.
}
\end{figure*}
It follows from the identity $\bm w^\pm\times \bm v=\mp \bm v\times \bm b$ that 
\begin{equation}
w^+\sin\theta^+=v\sin \theta = w^-\sin\theta^-, 
\end{equation}
where $\theta^+$ is the angle formed by $\bm w^+$ and $\bm v$, $\theta^-$ is the angle formed by 
$\bm w^-$ and $\bm v$, and $\theta$ is the angle formed by $\bm v$ and $\bm b$.
In addition, $\theta^+ +\theta^-=\pi/2$.  Following \citet{Boldyrev:2006}, suppose that the 
gradient of the fluctuations is in the direction perpendicular to $\bm v$.  In this case, 
\begin{equation}
(\bm w^- \cdot \nabla) \simeq \frac{w^- \sin\theta^-}{\lambda_\perp}
=\frac{v \sin\theta}{\lambda_\perp}
\label{NL_term1}
\end{equation}
and
\begin{equation}
(\bm w^+ \cdot \nabla) \simeq \frac{w^+ \sin\theta^+}{\lambda_\perp}
=\frac{v \sin\theta}{\lambda_\perp}.
\label{NL_term2}
\end{equation}
The time rate of change caused by nonlinear interactions is estimated 
from the relations
\begin{equation}
\frac{\partial}{\partial t} \frac{|\bm w^+|^2}{2} \simeq \bm w^+ \cdot 
(\bm w^- \cdot \nabla)\bm w^+ 
\label{NL_terms1}
\end{equation}
and 
\begin{equation}
\frac{\partial}{\partial t} \frac{|\bm w^-|^2}{2} \simeq \bm w^- \cdot 
(\bm w^+ \cdot \nabla)\bm w^-. 
\label{NL_terms2}
\end{equation}

If $\bm v$ and $\bm b$ are aligned with some small angle $\theta$, then the fluctuations
are called ``positively aligned'' and denoted by `$\uparrow$' (Figure 1a).  
Similarly, if $\bm v$ and $-\bm b$ are aligned with some small angle $\theta$, then the 
fluctuations are called ``negatively aligned'' or ``anti-aligned'' and denoted by 
`$\downarrow$' (Figure 1b).  For positively aligned fluctuations, equations 
(\ref{NL_term1})--(\ref{NL_terms2}) imply
\begin{equation}
\frac{\partial}{\partial t} \frac{(w^+_\uparrow)^2}{2} \simeq  \frac{(w^+_\uparrow)^2 w^-_\uparrow 
\sin\theta^-_\uparrow}{\lambda_\perp} =  \frac{(w^+_\uparrow)^2 v_\uparrow
\sin\theta_\uparrow}{\lambda_\perp}   \label{w+up} 
\end{equation}
and
\begin{equation}
\frac{\partial}{\partial t} \frac{(w^-_\uparrow)^2}{2} \simeq  \frac{(w^-_\uparrow)^2 w^+_\uparrow 
\sin\theta^+_\uparrow}{\lambda_\perp} =  \frac{(w^-_\uparrow)^2 v_\uparrow
\sin\theta_\uparrow}{\lambda_\perp}, \label{w-up} 
\end{equation}
where $\theta_\uparrow$ is the angle formed by $\bm v$ and $\bm b$ and quantities
with the subscript $\uparrow$ describe positively aligned fluctuations.
It is clear from the middle term in equation (\ref{w+up}) that the time rate of change
of $w^+_\uparrow$ depends on $w^-_\uparrow$, consistent with the nonlinear terms in the
MHD equations, although this dependence is not immediately apparent in the last term 
in (\ref{w+up}).  For negatively aligned fluctuations, equations 
(\ref{NL_term1})--(\ref{NL_terms2}) imply
\begin{equation}
\frac{\partial}{\partial t} \frac{(w^+_\downarrow)^2}{2} \simeq  \frac{(w^+_\downarrow)^2 w^-_\downarrow 
\sin\theta^-_\downarrow}{\lambda_\perp} =  \frac{(w^+_\downarrow)^2 v_\downarrow
\sin\theta_\downarrow}{\lambda_\perp}   \label{w+down} 
\end{equation}
and
\begin{equation}
\frac{\partial}{\partial t} \frac{(w^-_\downarrow)^2}{2} \simeq  \frac{(w^-_\downarrow)^2 w^+_\downarrow 
\sin\theta^+_\downarrow}{\lambda_\perp} =  \frac{(w^-_\downarrow)^2 v_\downarrow
\sin\theta_\downarrow}{\lambda_\perp}, \label{w-down} 
\end{equation}
where $\theta_\downarrow$ is the angle formed by $\bm v$ and $-\bm b$ and quantities
with the subscript $\downarrow$ describe negatively aligned fluctuations.
Here, $0<\theta_\uparrow<\pi/2$ and $0<\theta_\downarrow<\pi/2$.

In general, the fluctuations $\bm v$ and $\bm b$ observed at any point $(\bm x,t)$ 
are either positively or negatively aligned.
For a point $(\bm x,t)$ picked at random, let $p$ and $q$ be the probabilities 
the alignment is positive or negative, respectively ($p+q=1$).  Then, on average,
\begin{equation}
\frac{\partial}{\partial t} \frac{(\tilde w^+)^2}{2} \simeq  \frac{1}
{\lambda_\perp}\big[ p(w^+_\uparrow)^2 v_\uparrow\sin\theta_\uparrow + q (w^+_\downarrow)^2 
v_\downarrow\sin\theta_\downarrow\big]   \label{NL1}   
\end{equation}
and 
\begin{equation}
\frac{\partial}{\partial t} \frac{(\tilde w^-)^2}{2} \simeq  \frac{1}
{\lambda_\perp}\big[ p(w^-_\uparrow)^2 v_\uparrow\sin\theta_\uparrow + q (w^-_\downarrow)^2 
v_\downarrow\sin\theta_\downarrow\big],   \label{NL2}   
\end{equation}
where the rms values $\tilde w^\pm$ are defined by
\begin{align}
(\tilde w^+)^2 &= p(w_\uparrow^+)^2+q(w_\downarrow^+)^2, \label{rms1} \\
(\tilde w^-)^2 &= p(w_\uparrow^-)^2+q(w_\downarrow^-)^2.  \label{rms2}
\end{align}
The following relations also hold.  For a positively aligned fluctuation, 
assuming $|\bm v|=|\bm b|$, 
\begin{align}
\bm w_\uparrow^+\cdot \bm w_\uparrow^+ &= 2v_\uparrow^2 (1+\cos\theta_\uparrow), \label{w_up} \\
\bm w_\uparrow^-\cdot \bm w_\uparrow^- &= 2v_\uparrow^2 (1-\cos\theta_\uparrow),
\end{align}
and $w^+_\uparrow w^-_\uparrow=2v_\uparrow^2\sin\theta_\uparrow$.  The energy of a positively 
aligned fluctuation is $v_\uparrow^2$.  For a negatively aligned fluctuation
\begin{align}
\bm w_\downarrow^+\cdot \bm w_\downarrow^+ &= 2v_\downarrow^2 (1-\cos\theta_\downarrow), \\
\bm w_\downarrow^-\cdot \bm w_\downarrow^- &= 2v_\downarrow^2 (1+\cos\theta_\downarrow),  \label{w_down} 
\end{align}
and $w^+_\downarrow w^-_\downarrow=2v_\downarrow^2\sin\theta_\downarrow$.
The energy of a negatively aligned fluctuation is $v_\downarrow^2$.  Thus, the rms values 
(\ref{rms1}) and (\ref{rms2}) are
\begin{align}
(\tilde w^+)^2 &= 2 \big[pv_\uparrow^2(1+\cos\theta_\uparrow)+qv_\downarrow^2(1-\cos\theta_\downarrow)\big], \label{w+} \\
(\tilde w^-)^2 &= 2 \big[pv_\uparrow^2(1-\cos\theta_\uparrow)+qv_\downarrow^2(1+\cos\theta_\downarrow)\big]. \label{w-}
\end{align}

If the angles are small, $\theta_\uparrow \ll 1$ and $\theta_\downarrow \ll 1$, 
then the small parameter $\theta$ can be used to order the
terms in equations (\ref{w+}) and (\ref{w-}) so that to leading order
\begin{equation}
(\tilde w^+)^2 \simeq 4v_\uparrow^2 p \qquad \mbox{and} \qquad (\tilde w^-)^2 \simeq 4v_\downarrow^2 q,
\label{partition}
\end{equation}
where $p+q=1$.  
This may be derived as follows. In equations (\ref{w+}) and (\ref{w-}) assume that the angles 
are both small and then substitute $1+\cos\theta \simeq 2$ and
$1-\cos\theta = 2\sin^2(\theta/2)$ to obtain 
\begin{equation}
(\tilde w^+)^2 \simeq 4 \big[pv_\uparrow^2+qv_\downarrow^2 \sin^2(\theta_\downarrow/2)\big]  \\
\end{equation}
and
\begin{equation}
(\tilde w^-)^2 \simeq 4 \big[pv_\uparrow^2 \sin^2(\theta_\uparrow/2)+qv_\downarrow^2\big]. 
\end{equation}
As $\lambda_\perp \rightarrow 0$, both $\theta_\uparrow \rightarrow 0$ and 
$\theta_\downarrow \rightarrow 0$ and, therefore, to first order, the
terms proportional to $\sin^2(\theta)$ may be neglected.  Alternatively, note that 
\begin{equation}
\bigg(\frac{\tilde w^+}{\tilde w^-}\bigg)^2 \simeq \frac{(pv_\uparrow^2/qv_\downarrow^2)
+ \sin^2(\theta_\downarrow/2)}{(pv_\uparrow^2/qv_\downarrow^2)\sin^2(\theta_\uparrow/2)+1}. 
\end{equation}
As discussed below, solar wind observations show that this quantity is approximately 
constant in the inertial range.
Now, as $\theta_\uparrow \rightarrow 0$ and 
$\theta_\downarrow \rightarrow 0$ the only way that this can remain constant is
if $pv_\uparrow^2/qv_\downarrow^2$ is bounded away from zero and 
\begin{equation}
\bigg(\frac{\tilde w^+}{\tilde w^-}\bigg)^2 \simeq \frac{pv_\uparrow^2}{qv_\downarrow^2}.
\label{ratio}
\end{equation}
This justifies the approximation in Eqn (\ref{partition}).

Equation (\ref{partition}) shows that at a given scale $\lambda_\perp$ the total 
energy $[(\tilde w^+)^2 +(\tilde w^-)^2]/4$
is partitioned into two parts, the energy $(\tilde w^+)^2/4$  associated with 
positive alignment and the energy $(\tilde w^-)^2/4$  associated with negative 
alignment.  The normalized cross-helicity $\sigma_c$ is defined as the ratio of the
cross-helicity to the energy at a given scale and can be written
\begin{equation}
\sigma_c = \frac{(\tilde w^+)^2-(\tilde w^-)^2}{(\tilde w^+)^2+(\tilde w^-)^2}.   \label{sigma}
\end{equation}

For small angles, equations (\ref{NL1}) and (\ref{NL2}) become, to leading order,
\begin{align}
\frac{\partial}{\partial t} \frac{(\tilde w^+)^2}{2} &\simeq  \frac{4pv_\uparrow^3 \theta_\uparrow}{\lambda_\perp},   \label{NL1b}   \\
\frac{\partial}{\partial t} \frac{(\tilde w^-)^2}{2} &\simeq  \frac{4qv_\downarrow^3 \theta_\downarrow}{\lambda_\perp}.    \label{NL2b}   
\end{align}
To express these in terms of the rms values $\tilde w^\pm$, eliminate $v_\uparrow$ and
$v_\downarrow$ using equation (\ref{partition}).  This yields
\begin{align}
\frac{\partial}{\partial t} \frac{(\tilde w^+)^2}{2} &\simeq  \frac{(\tilde w^+)^3 
\theta_\uparrow} {2\lambda_\perp p^{1/2}},   \label{NL1c}   \\
\frac{\partial}{\partial t} \frac{(\tilde w^-)^2}{2} &\simeq  \frac{(\tilde w^-)^3 
\theta_\downarrow} {2\lambda_\perp q^{1/2}}.     \label{NL2c}   
\end{align}
These estimates shall be used to derive the cascade times.  


\section{Energy cascade time}

When nonlinear interactions are strong and a large number of Fourier modes are excited, 
fluctuations occur continuously in time and space.  During a time $\tau$ the fractional 
change in the quantity $(\tilde w^+)^2$ is, from (\ref{NL1c}),
\begin{equation}
\chi^+(\tau) \simeq \frac{(w^+)^3 \theta_\uparrow} {2\lambda_\perp p^{1/2}}\cdot 
\frac{2\tau}{(w^+)^2}=\frac{w^+\theta_\uparrow \tau}{\lambda_\perp p^{1/2}}, \qquad \tau\le \tau^+,
\label{chi+}
\end{equation}
where $\tau^+$ is the cascade time at the lengthscale $\lambda_\perp$ and
the tildes have been dropped.  Similarly, the fractional change in the quantity $(\tilde w^-)^2$ 
is, from (\ref{NL2c}),
\begin{equation}
\chi^-(\tau) \simeq \frac{(w^-)^3 \theta_\downarrow} {2\lambda_\perp q^{1/2}}\cdot 
\frac{2\tau}{(w^-)^2}=\frac{w^-\theta_\downarrow \tau}{\lambda_\perp q^{1/2}}, \qquad \tau\le \tau^-,
\label{chi-}
\end{equation}
where $\tau^-$ is the cascade time of $\tilde w^-$ and the tildes have been dropped for brevity.
Hereafter, the tildes will be omitted and $w^+$ and $w^-$ will always represent the rms values.  

According to the definition of the energy cascade time, the fractional change $\chi^+$ is 
of order unity when the interaction time $\tau$ is equal to the cascade time $\tau^+$.  
Therefore, the relations (\ref{chi+}) and (\ref{chi-}) imply
\begin{equation}
\tau^+ \simeq \frac{\lambda_\perp p^{1/2}}{w^+\theta_\uparrow}, \qquad 
\tau^- \simeq \frac{\lambda_\perp q^{1/2}}{w^-\theta_\downarrow}.
\label{t_cas}
\end{equation}
By similar reasoning, equations (\ref{w+up}) and (\ref{w-down}) imply
\begin{equation}
\tau_\uparrow \simeq \frac{\lambda_\perp}{2v_\uparrow \theta_\uparrow}, \qquad 
\tau_\downarrow \simeq \frac{\lambda_\perp}{2v_\downarrow \theta_\downarrow}.
\label{t_up_down}
\end{equation}
Moreover, equations (\ref{t_cas}), (\ref{t_up_down}), and (\ref{partition}) imply 
$\tau^+ = \tau_\uparrow$ and $\tau^-= \tau_\downarrow$.  Thus, the
energy cascade times for the rms Elsasser amplitudes are equal to the
energy cascade times for the positively and negatively aligned fluctuations.

For balanced turbulence, $\sigma_c\rightarrow 0$, $w^+/w^-\rightarrow 1$, $p=q$, 
$\theta_\uparrow=\theta_\downarrow$,
and the energy cascade times (\ref{t_cas}) reduce to the cascade time in Boldyrev's original
theory \citep{Boldyrev:2006}.  For imbalanced turbulence, $\sigma_c\ne 0$, 
the cascade times (\ref{t_cas}) are different from the cascade times
$\tau^\pm \sim \lambda_\perp/w^\mp \theta^\mp$ in the theory of 
Perez \& Boldyrev \citep{Perez:2009}.  
The theory presented here is different from the theory of Perez \& Boldyrev \citep{Perez:2009}
because the latter theory does not take into account the existence of two separate types of 
fluctuations, positively and negatively aligned, with separate probabilities of occurrence $p$ and $q$.
Taking this into account and also the 
definitions of the rms amplitudes (\ref{rms1}) and (\ref{rms2}), it follows from the
preceding analysis that the timescales for the rms amplitudes take the form (\ref{t_cas}).  


As pointed out by 
Kraichnan \citep{Kraichnan:1965}, Dobrowolny, Mangeney, and Veltri 
\citep{Dobrowolny_Mangeney:1980}, and others, the energy cascade in MHD turbulence 
occurs through collisions between Alfv\'en wavepackets propagating in
opposite directions along the mean magnetic field.  In other words, it is the
interaction between $w^+$ and $w^-$ waves that causes the energy to cascade to smaller
scales in MHD turbulence.  Consequently, the cascade time for $w^+$, say, 
should depend on $w^-$.  While it may appear from equations 
(\ref{t_cas})--(\ref{t_up_down}) that the timescale for $w^+$ 
fluctuations depends only on $w^+$ and, therefore, the interaction with
the $w^-$ waves is absent, this is not true.  The interactions are still present in
the expressions (\ref{t_cas}) and (\ref{t_up_down}) through the dependence on the angles 
and other parameters as will be shown in the next section.

\section{Theory of the energy cascade process}

Assuming there is no
direct injection of energy or cross-helicity within the inertial range
and there is no dissipation of energy or cross-helicity within the inertial range,
the energy cascade rate $\eps$ and the cross-helicity cascade rate $\eps_c$ are 
scale-invariant in the inertial range. 
It follows that the energy cascade rates for the two Elsasser variables $\eps^\pm = \eps
\pm \eps_c$ are also scale-invariant.  The theory of the energy cascade process 
is based on Kolmogorov's relations
\begin{equation}
\frac{(w^+)^2}{2\tau^+} = \eps^+ \qquad \mbox{and} \qquad 
\frac{(w^-)^2}{2\tau^-} = \eps^-,
\label{Kol}
\end{equation}
where the non-zero constants $\eps^+$ and $\eps^-$ are the energy cascade rates
per unit mass for the two Elsasser variables $w^+$ and $w^-$, respectively.  
These equations describe the conservation of energy
flux in $\bm k$-space (Fourier space).  In addition to Kolmogorov's relations,
there are two observational constraints that must be taken into account.
\medskip

Solar wind observations show that the energy and cross-helicity
spectra of the turbulence follow approximately the same power law 
in the inertial range (Fig.~2) 
\begin{figure}
\includegraphics[width=2.8in]{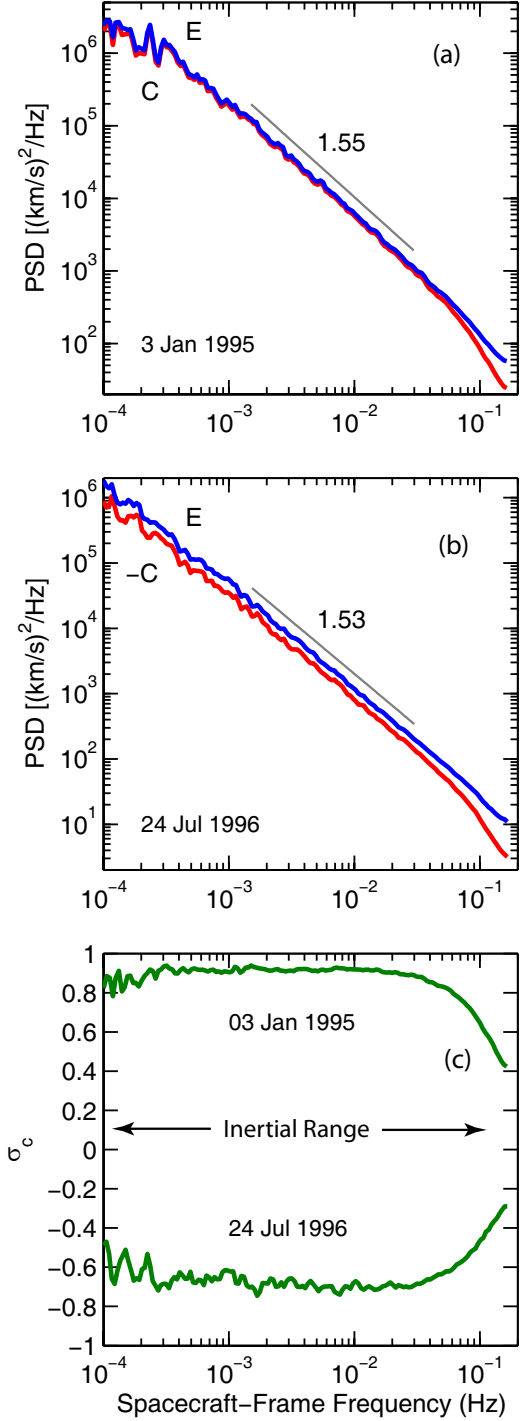}%
\caption{\label{fig1}%
     Typical energy $E$ and cross-helicity spectra $C$ (trace spectra) obtained using
3-second plasma velocity and magnetic field data from the {\it Wind}
spacecraft near the orbit of the earth at 1 AU.
(a) An interval of highly Alfv\'enic high-speed wind from 3 Jan 1995 09:00 to 8 Jan 1995 00:00,  
4.625 days.  (b) A weak high-speed stream embedded in low-speed wind;
24 Jul 1996 12:00 to 7 Aug 1996 00:00, 14 days.  
(c) The normalized cross-helicity $\sigma_c=C/E$ as a function of frequency.
The rapid change in $\sigma_c$ near the Nyquist frequency is at least partly caused by 
the FFT processing techniques and may not be a real physical effect.
}
\end{figure}
which implies that the normalized cross-helicity
$\sigma_c$ is approximately constant.  
In other words, the quantity $\sigma_c$ is approximately 
scale invariant.  Similar results have been found in simulations of incompressible MHD 
turbulence \citep{Verma_Roberts:1996,Perez:2009,Beresnyak_Lazarian:2008}.  
In particular, the 3D simulations of Perez \& Boldyrev \citep{Perez:2009} indicate
that the perpendicular Elsasser spectra are proportional to each other in Fourier space.
Solar wind observations also suggest that the probabilities $p$ and $q$ are 
approximately scale invariant as shown in Fig.~3.  These observations will now be 
taken into account in the theory.
\medskip

Assuming  $\sigma_c$ and $p$ are both scale invariant quantities, then $w^+/w^-$, 
$v_\uparrow/v_\downarrow$, $\tau^+/\tau^-$, and $\theta_\downarrow/\theta_\uparrow$ 
are scale invariant by equations (\ref{sigma}), (\ref{ratio}), (\ref{Kol}), 
and (\ref{t_cas}), respectively. 
In all, there are six different scale invariant ratios in the theory 
\begin{equation}
\frac{w^+}{w^-}, \quad \frac{\eps^+}{\eps^-}, \quad
\frac{p}{q}, \quad \frac{v_\uparrow}{v_\downarrow}, \quad \frac{\tau^+}{\tau^-}, \quad   
\frac{\theta_\uparrow}{\theta_\downarrow}. 
\label{six}
\end{equation}
At most, only three of these are independent, say, the first three.  
Equations (\ref{ratio}), (\ref{Kol}), and (\ref{t_cas}) imply
\begin{gather}
\frac{v_\uparrow}{v_\downarrow} = \sqrt{\frac{q}{p}} \cdot \frac{w^+}{w^-},  \\
\frac{\tau^+}{\tau^-}= \left(\frac{w^+}{w^-}\right)^2\frac{\eps^-}{\eps^+},  \label{ratio_tau} \\
\frac{\theta_\downarrow}{\theta_\uparrow} = \sqrt{\frac{q}{p}} 
\left(\frac{w^+}{w^-}\right)^3\frac{\eps^-}{\eps^+}.  \label{ratio_theta}
\end{gather}
Therefore, the six scale invariant ratios (\ref{six}) can all be 
expressed in terms of the first three $w^+/w^-$, $\eps^+/\eps^-$, and $p/q$.
\begin{figure}
\begin{center}
\includegraphics[width=3in]{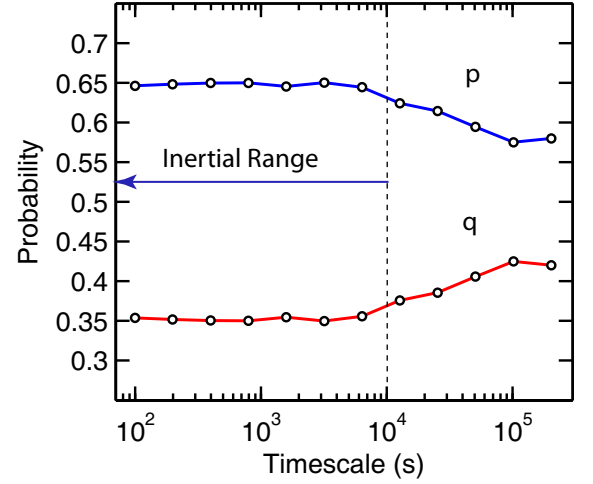}%
\caption{\label{fig2}%
     The probabilities $p$ and $q$ obtained from solar wind data by integrating
the observed probability density function for the angle $\theta$ from 0 to $\pi/2$ and
from $\pi/2$ to $\pi$, respectively.  The data was acquired by the {\it Wind} spacecraft between
8 Jan 1997 and 9 June 1997 and analyzed using the techniques described in 
\citep{Podesta_Chandran:2009}.  Examples of the probability density functions
can be found in \citep{Podesta_Chandran:2009}.
}
\end{center}
\end{figure}
\medskip

To be able to solve Kolmogorov's relations (\ref{Kol}) for $w^\pm$ it is necessary to express 
the alignment angle $\theta_\uparrow$ in terms of $w^\pm$.  In general, $\theta_\uparrow$ can 
depend on $w^+$, $w^-$, the Alfv\'en speed $v_A$, the lengthscale $\lambda_\perp$,
the cascade rates $\eps^+$ and $\eps^-$, and the probabilities $p$ and $q$.
By dimensional analysis,  $\theta_\uparrow$ must be a function of the following 
six dimensionless quantities
\begin{equation}
\frac{w^+}{v_A}, \quad \frac{w^-}{v_A}, \quad \frac{\eps^+ \lambda_\perp}{v_A^3}, \quad
\frac{\eps^- \lambda_\perp}{v_A^3}, \quad p, \quad q.  
\end{equation}
Moreover, $\theta_\uparrow$ must change into $\theta_\downarrow$
when $w^+$, $\eps^+$, and $p$ are interchanged with $w^-$, $\eps^-$, and $q$, respectively,
to be consistent with the nonlinear terms (\ref{NL1c}) and (\ref{NL2c}).  For a theory 
composed of power law functions, the only forms that satisfy all these requirements are
\begin{align}
\theta_\uparrow &\propto \bigg(\frac{w^+}{v_A}\bigg)^{\!\alpha}
\bigg(\frac{w^-}{v_A}\bigg)^{\!\beta}
\bigg(\frac{\eps^+ \lambda_\perp}{v_A^3}\bigg)^{\!\gamma}
\bigg(\frac{\eps^- \lambda_\perp}{v_A^3}\bigg)^{\!\delta}
p^\mu q^\nu, 
\label{theta_up}  \\
\theta_\downarrow &\propto \bigg(\frac{w^-}{v_A}\bigg)^{\!\alpha}
\bigg(\frac{w^+}{v_A}\bigg)^{\!\beta}
\bigg(\frac{\eps^- \lambda_\perp}{v_A^3}\bigg)^{\!\gamma}
\bigg(\frac{\eps^+ \lambda_\perp}{v_A^3}\bigg)^{\!\delta}
q^\mu p^\nu,
\label{theta_down}
\end{align}
where $\alpha$, $\beta$, $\gamma$, $\delta$, $\mu$, and $\nu$ are constants that must be 
determined by the theory.  In addition, there is a leading coefficient which is omitted.
\medskip

The substitution of (\ref{theta_up}) and (\ref{theta_down}) into equation 
(\ref{ratio_theta}) yields $\beta=\alpha+3$, $\delta=\gamma -1$, and $\nu=\mu-1/2$.  
The parameters are further constrained by 
considering the geometry of the ``turbulent eddies'' associated with the fluctuations 
$v_\uparrow$ and $v_\downarrow$.  The parallel correlation length is defined by 
$\lambda_\parallel^\uparrow=v_A\tau_\uparrow$ and the correlation length in the direction
of the velocity fluctuation is $\bm \xi_\uparrow=\bm v_\uparrow\tau_\uparrow$.  Similarly, 
the correlation lengths for negatively aligned fluctuations are 
$\lambda_\parallel^\downarrow=v_A\tau_\downarrow$ 
and $\xi_\downarrow=v_\downarrow\tau_\downarrow$. In the plane 
perpendicular to the local mean magnetic field $\bm \xi$ is parallel to $\bm v$,
the gradient direction is perpendicular to $\bm v$ with lengthscale $\lambda_\perp$, 
and the eddy dimensions are $\xi \times \lambda_\perp$.  The dimension parallel
to the mean magnetic field is $\lambda_\parallel$.  Hence, in physical space the
turbulent eddies can be visualized as three-dimensional structures with dimensions 
$\lambda_\perp \times \xi \times \lambda_\parallel$.   
\medskip

The coherence times for longitudinal and transverse motions of the eddy must be equal to each
other and also to the cascade time.  This is the {\it critical balance} condition
of Goldreich and Sridhar  
which is also implicit in the work of Higdon \cite{Higdon:1984}.
Equation (\ref{t_up_down}) and the definitions of the correlation lengths 
in the last paragraph immediately yield
the critical balance condition
\begin{equation}
\tau_\uparrow = \frac{\lambda_\parallel^\uparrow}{v_A} = \frac{\xi_\uparrow}{v_\uparrow}\simeq  
\frac{\lambda_\perp}{2v_\uparrow\theta_{\uparrow}} 
\label{cb1} 
\end{equation}
with a similar condition for the negatively aligned fluctuations
\begin{equation}
\tau_\downarrow = \frac{\lambda_\parallel^\downarrow}{v_A} = \frac{\xi_\downarrow}{v_\downarrow}\simeq  
\frac{\lambda_\perp}{2v_\downarrow\theta_{\downarrow}}. 
\end{equation}

Now consider the eddy geometry.  When the mean magnetic field is strong enough that 
$w^\pm/v_A<1$, then $\lambda_\parallel > \xi > \lambda_\perp$ and the eddies are elongated
in the parallel direction.  The condition $w^\pm/v_A<1$ is assumed hereafter.
Equation (\ref{cb1}) shows that the aspect ratio in the field 
perpendicular plane is $\phi_\uparrow = \lambda_\perp/\xi_\uparrow =2\theta_\uparrow$ 
and the aspect ratio in the parallel 
direction is, from equations (\ref{cb1}) and (\ref{partition}), 
\begin{equation}
\psi_\uparrow = \frac{\xi_\uparrow}{\lambda_\parallel^\uparrow} = \frac{v_\uparrow}{v_A} =
\frac{w^+}{2v_A p^{1/2}}. 
\end{equation}
The two aspect ratios will scale in the same way if $\phi_\uparrow/\psi_\uparrow$ is scale 
invariant.  This implies that $\alpha=-1$ and $\gamma=1/2$.  
The assumption that the ratio $\phi_\uparrow/\psi_\uparrow$ is scale 
invariant is different from Boldyrev's original approach in which he assumed
that the alignment angles in and out of the field perpendicular plane
are simultaneously minimized.
Nevertheless, our assumption retains the spirit of Boldyrev's original theory 
which implies the geometry of turbulent fluctuations are scale-invariant.
\medskip

Solving Kolmogorov's relation (\ref{Kol}) using (\ref{t_cas}), (\ref{theta_up}), 
(\ref{theta_down}), and the parameter values 
obtained so far, one finds 
\begin{equation}
\frac{w^\pm}{v_A} \simeq  \bigg(\frac{w^+}{w^-}\bigg)^{\! \pm 1/2}
\bigg(\frac{\eps^-}{\eps^+}\bigg)^{\! \pm 1/8}
\bigg(\frac{\eps^\pm\lambda_\perp}{v_A^3}\bigg)^{\!1/4} (pq)^{-\nu/4}  \label{W+} 
\end{equation}
and the total energy cascade rate $\eps=(\eps^+ + \eps^-)/2$ is 
\begin{equation}
\eps= \frac{(w^+w^-)^2}{4v_A \lambda_\perp} \Bigg( \sqrt{\frac{\eps^+}{\eps^-}}
+\sqrt{\frac{\eps^-}{\eps^+}}\Bigg) (pq)^{\nu}. 
\end{equation}
The total energy at scale $\lambda_\perp$ is 
\begin{equation}
\frac{(w^+)^2+(w^-)^2}{4}=\frac{w^+w^-}{4} \bigg(\frac{w^+}{w^-}+\frac{w^-}{w^+}\bigg)\equiv v^2.
\end{equation}
Therefore, the energy cascade rate can be written
\begin{equation}
\epsilon= \frac{4v^4}{v_A \lambda_\perp} \Bigg(\sqrt{\frac{\eps^+}{\eps^-}}
+\sqrt{\frac{\eps^-}{\eps^+}}\Bigg) 
\bigg(\frac{w^+}{w^-}+\frac{w^-}{w^+}\bigg)^{\!-2}(pq)^{\nu}. 
\label{eps}
\end{equation}
Assuming the rms energy $v^2$ at scale $\lambda_\perp$ is 
held constant, the terms on the right-hand side describe the
dependence of the energy cascade rate on the ratios $\eps^+/\eps^-$ and $w^+/w^-$.  
The value of $\nu$ may be determined by comparison with 
experiment or possibly by further physical considerations.  This parameter does not 
affect the inertial range scaling laws and is left undetermined for the moment.  

At this point it is of interest to return to the expressions (\ref{t_cas}) for the 
cascade times and ask: How do the cascade times depend on the rms
Elsasser amplitudes?  Using the parameter values obtained previously, equation
(\ref{theta_up}) becomes 
\begin{equation}
\theta_\uparrow \sim \bigg(\frac{w^+}{v_A}\bigg)^{\!-1}\bigg(\frac{w^-}{v_A}\bigg)^{\!2}
\bigg(\frac{\eps^+ \lambda_\perp}{v_A^3}\bigg)^{\!1/2}
\bigg(\frac{\eps^- \lambda_\perp}{v_A^3}\bigg)^{\!-1/2}
p^{\nu+1/2} q^\nu 
\end{equation}
and the substitution of this result into equation (\ref{t_cas}) yields
\begin{equation}
\tau^+ \simeq \frac{\lambda_\perp }{v_A} 
\bigg(\frac{v_A}{w^-}\bigg)^{\!2}\bigg(\frac{\eps^-}{\eps^+}\bigg)^{\! 1/2}
(pq)^{-\nu}.
\label{t_cas_full}
\end{equation}
A similar expression holds for $\tau^-$ so that the ratio $\tau^+/\tau^-$ satisfies 
(\ref{ratio_tau}). Ignoring scale invariant factors, the preceding equation shows that
\begin{equation}
\tau^+ \propto \frac{\lambda_\perp v_A}{(w^-)^2} \qquad \mbox{and} \qquad
\tau^- \propto \frac{\lambda_\perp v_A}{(w^+)^2}.
\label{t_cascade}
\end{equation}
In this form, the angle dependence has been eliminated.  Note that the simple
estimate $\tau^+ \sim \lambda_\perp/w^-$ suggested by the nonlinear term in the MHD
equations is modified by the factor $v_A/w^-$ which accounts for the 
weakening of nonlinear interactions caused by scale dependent alignment.
The presence of this algebraic factor is one of the hallmarks of Boldyrev's original 
(2006) theory which is generalized here to imbalanced turbulence.  Remarkably,
the relations (\ref{t_cascade}) are identical to those in the isotropic theory of imbalanced 
turbulence developed by Dobrowolny, Mangeney, and Veltri; see equation (10) in 
\citep{Dobrowolny_Mangeney:1980}.  Recall that Dobrowolny, Mangeney, and Veltri concluded
from their expressions for the cascade times that {\it steady state} turbulence with 
nonvanishing cross-helicity is impossible. On the contrary, the theory presented 
here allows such a steady state because the additional coefficients shown in 
(\ref{t_cas_full}) but not (\ref{t_cascade}) maintain the
relation (\ref{ratio_tau}) even when $\eps^+\ne \eps^-$.  Thus, the theory presented 
here is also a generalization of the theory of Dobrowolny, Mangeney, and Veltri
\citep{Dobrowolny_Mangeney:1980}.

A remark about the timescales in the theory should be mentioned.  If $w^+>w^-$, then 
equation (\ref{ratio_tau}) implies it is possible that $\tau^+<\tau^-$
since there is nothing in the theory that prevents this.
That is, the energy of the more energetic Elsasser species may be transferred to
smaller scales in less time than the energy of the less energetic Elsasser species.
This is not inconsistent with dynamic alignment, a well known
effect seen in simulations of decaying incompressible MHD turbulence where the
minority species usually decays more rapidly than the dominant species causing the magnitude
of the normalized cross-helicity to increase with time \citep{Dobrowolny_Mangeney:1980,
Matthaeus_Goldstein:1983,Matthaeus_Montgomery:1984,Pouquet_Frisch:1986}.  
\medskip

In freely decaying turbulence, dynamic alignment occurs whenever the total energy decays 
more rapidly than the cross-helicity, that is, $\eps >|\eps_c|$, where the
cascade rate of cross-helicity $\eps_c$ may be positive or negative.  From the relations 
$\eps > 0$ and $\eps^\pm =\eps \pm \eps_c$, it follows 
that dynamic alignment occurs if and only if $\eps^+>0$ and $\eps^->0$.  If $w^+> w^-$, 
it is not necessary that $\tau^+>\tau^-$, only that
\begin{equation}
\frac{\tau^+}{\tau^-}>  \frac{\eps^-}{\eps^+},
\label{ratio_tau1}
\end{equation}
as can be seen from equation (\ref{ratio_tau}).
Therefore, even though the relation $\tau^+<\tau^-$ may seem counter-intuitive,
it is not inconsistent with dynamic alignment.  
\medskip


\section{Summary and Conclusions}

Observations of scale dependent alignment of velocity and magnetic field fluctuations
$\delta \bm v_\perp$ and $\delta \bm b_\perp$ in the solar wind suggest that this 
effect must be included in any theory of solar wind turbulence 
\citep{Podesta_Bhattacharjee:2008,Podesta_Chandran:2009}.
\citet{Perez:2009} have recently discussed a theory of imbalanced turbulence that
includes scale dependent alignment of the fluctuations $\delta \bm v_\perp$ and 
$\delta \bm b_\perp$ in the inertial range.  We have extended the 
Perez-Boldyrev theory by including the probabilities $p$ and $q$
which solar wind observations indicate are not necessarily equal.  
Operationally, the probabilities $p$ and $q$ may be defined as follows.
Suppose space is covered by a uniform cartesian grid or three dimensional mesh.
At each grid-point one may compute the fluctuations $\delta \bm v_\perp$ and 
$\delta \bm b_\perp$ and the angle between them $\theta$.
If the angle lies in the range $0<\theta<\pi/2$, then the fluctuation
is positively aligned and if  $\pi/2<\theta<\pi$, then the fluctuation
is negatively aligned.  By counting the number of positively and negatively aligned
fluctuations in a large volume $V$, much larger than the lengthscales of the
turbulent eddies, the probabilities $p$ and $q$ may be defined as
the fractional numbers of positively and negatively aligned fluctuations
in the volume $V$.  
\medskip

The phenomenological theory developed in this paper was guided primarily by 
two new solar wind observations.
It should be noted that both of these solar wind observations are 
necessary for the development of the theory.
At first glance, it may seem that the condition $\sigma_c=$ {\it const} implies that 
$p$ and $q$ are both constant.  Or that these two conditions are somehow
equivalent.  However, the relation 
$(w^+/w^-)^2\simeq pv_\uparrow^2/qv_\downarrow^2$, equation (\ref{ratio}),  shows that
$p/q$ can vary with the lengthscale even if $w^+/w^-$ is constant.  Therefore, 
it is essential to have separate observations of the scale invariance of $\sigma_c$ 
and the scale invariance of $p$ and $q$ to support the theoretical
framework developed here.
\medskip

In summary, using estimates of the cascade times derived from the nonlinear terms in
the incompressible MHD equations and two new observational constraints derived 
from studies of solar wind data, we have constructed a generalization of Boldyrev's 
theory \citep{Boldyrev:2006}
that depends on the three parameters $w^+/w^-$, $\eps^+/\eps^-$, and $p/q$.
The theory reduces to the original theory of 
\citet{Boldyrev:2006} when $w^+ = w^-$, $\epsilon^+=\epsilon^-$, and $p=q$ since 
in this limit $\theta_\uparrow =\theta_\downarrow$
and the cascade times (\ref{t_cas}) become equal to those of \citet{Boldyrev:2006}. 
For imbalanced turbulence $w^+\ne w^-$, $p\ne q$, and the theory predicts the
scaling laws $w^\pm \propto \lambda_\perp^{1/4}$, $\theta_{\uparrow \downarrow}\propto 
\lambda_\perp^{1/4}$, and $\lambda_\parallel^\pm \propto \lambda_\perp^{1/2}$. 
Interestingly, the scaling laws for balanced and imbalanced turbulence are
the same.
The perpendicular energy spectrum defined by  $k_\perp E^\pm \sim |w^\pm|^2$  
has the inertial range scaling $E^\pm \propto k_\perp^{-3/2}$ with 
\begin{equation}
\frac{E^+}{E^-}= \bigg(\frac{w^+}{w^-}\bigg)^{\!2} = \frac{1+\sigma_c}{1-\sigma_c} =\mbox{const}. 
\label{relation2}
\end{equation}
The theory assumes that the cascades for positively and negatively aligned fluctuations 
are both in a state of critical balance (\ref{cb1}), although they are governed by 
different timescales, and that the eddy geometry is scale invariant.  
The positively aligned fluctuations occupy a fractional
volume $p$ and the negatively aligned fluctuations occupy a fractional volume $q$ so that
the energy cascade rate is 
\begin{equation}
\eps = p\frac{v_\uparrow^2}{\tau_\uparrow}+q\frac{v_\downarrow^2}{\tau_\downarrow}
\end{equation}
or, equivalently, 
\begin{equation}
\eps = \frac{(w^+)^2}{4\tau^+}+\frac{(w^-)^2}{4\tau^-}.
\end{equation}
\medskip

In the discussion following equation (\ref{six}) it was shown that at most three 
of the ratios $w^+/w^-$, $\eps^+/\eps^-$, and $p/q$ can be independent.  However, 
the two ratios $w^+/w^-$ and $\eps^+/\eps^-$ cannot be independent since in the case 
of homogeneous steady-state turbulence $w^+ = w^-$ implies $\eps^+=\eps^-$ and 
vice versa.  This is because the injection of cross-helicity into the
system, $\eps_c \ne 0$ or $\eps^+ \ne \eps^-$, will create 
a nonzero cross-helicity spectrum and a cascade of cross-helicity from large to small 
scales which implies a net accumulation of cross-helicity within the volume ($\sigma_c\ne 0$).  
Hence, at most two of the ratios and $w^+/w^-$ and $p/q$ are independent.
Whether $p/q$ can be expressed in terms of $w^+/w^-$ and $\eps^+/\eps^-$
is an open question. 
\medskip

\begin{acknowledgements}

We are grateful to S.~Boldyrev for valuable comments on an earlier version of the
manuscript and to Pablo Mininni and Jean Perez for helpful discussions. This research 
is supported by DOE grant number DE-FG02-07ER46372, NASA grant number NNX06AC19G, and NSF. 
Additional support for John Podesta comes from the NASA Solar and Heliospheric Physics Program and the
NSF SHINE Program.

\end{acknowledgements}



\hyphenation{Post-Script Sprin-ger}

\end{document}